\documentclass[aps,pra,superscriptaddress,twocolumn,amsmath,amssymb]{revtex4-2}
\usepackage{amsmath,amsfonts,amssymb,graphicx}
\usepackage{epsfig}
\usepackage{epstopdf}
\newcommand{\be}{\begin{equation}}
\newcommand{\ee}{\end{equation}}
\newcommand{\bey}{\begin{eqnarray}}
\newcommand{\eey}{\end{eqnarray}}
\newcommand{\bw}{\begin{widetext}}
	\newcommand{\ew}{\end{widetext}}

\newcommand{\ba}{\begin{array}}
	\newcommand{\ea}{\end{array}}
\newcommand{\bi}{\begin{itemize}}
	\newcommand{\ei}{\end{itemize}}
\newcommand{\bem}{\begin{enumerate}}
	\newcommand{\eem}{\end{enumerate}}

\newcommand{\hefei}{Department of Modern Physics, University of Science and Technology of China, Hefei 230026, China}
\newcommand{\hefeiMMR}{CAS Key Laboratory of Microscale Magnetic Resonance,
	University of Science and Technology of China, Hefei 230026, China}
\newcommand{\como}{Center for Nonlinear and Complex Systems, Dipartimento di Scienza e Alta Tecnologia,
	Universit\`a degli Studi dell'Insubria, via Valleggio 11, 22100 Como, Italy}
\newcommand{\infn}{Istituto Nazionale di Fisica Nucleare, Sezione di Milano, via Celoria 16, 20133 Milano, Italy}
\newcommand{\brazil}{International Institute of Physics, Federal University of Rio Grande do Norte,
	Campus Universit\'ario - Lagoa Nova, CP. 1613, Natal, Rio Grande Do Norte 59078-970, Brazil}
\newcommand{\NEST}{NEST, Istituto Nanoscienze-CNR, I-56126 Pisa, Italy}

\begin{document}

\title{Statistical and dynamical properties of quantum triangle map}
	
\author{Jiaozi Wang}
\email{jiaozi.wang@uos.de}
\affiliation{Department of Physics, University of Osnabr\"uck, D-49069 
	Osnabr\"uck, Germany}
	\author{Giuliano Benenti}
	\email{giuliano.benenti@uninsubria.it}
	\affiliation{\como}
	\affiliation{\infn}
	\affiliation{\NEST}
	\author{Giulio Casati}
	\email{giulio.casati@uninsubria.it}
	\affiliation{\como}
	\affiliation{\brazil}
	\author{Wen-ge Wang}
	\email{wgwang@ustc.edu.cn}
	\affiliation{\hefei}
	\affiliation{\hefeiMMR}

	\date{\today}

	\begin{abstract}
		We study the statistical and dynamical properties of the quantum triangle map, 
		whose classical counterpart can exhibit ergodic and mixing dynamics, but is 
		never chaotic. Numerical results show that ergodicity is a sufficient condition 
		for spectrum and eigenfunctions to follow the prediction of 
		Random Matrix Theory, even though the underlying classical 
		dynamics is not chaotic. On the other hand, dynamical quantities such as the 
		out-of-time-ordered correlator (OTOC) and the number of harmonics, exhibit a 
		growth rate vanishing in the semiclassical limit, in agreement with the 
		fact that classical dynamics has zero Lyapunov exponent. 
		Our finding show that, while  spectral statistics can be used 
		to detect ergodicity, OTOC and number of harmonics are 
		diagnostics of chaos. 
	\end{abstract}
	\maketitle

\section{Introduction}

Classical ergodic theory provides a useful tool for the investigation of the statistical
properties of classical dynamical systems with finite number of degrees of freedom. 
Indeed, the classical motion exhibits a rich variety of properties, ranging from
complete integrability up to the exponential unstable chaotic motion, 
indistinguishable from a purely random process. The question if, and to what extent, 
this rich variety and beautiful complexity and intricacy of classical motion manifests
itself in the quantum domain, has been the subject of quantum chaos. 
Here we refer the reader to the beautiful book of Fritz Haake~\cite{Haake},
to whom this article is dedicated. 

In other words: in
the pure quantum context can we identify dynamical properties which play the same role 
as classical ergodic properties? In this connection a central role is played by 
the fact that the energy (or frequency) spectrum of any quantum system, 
bounded in phase space
and with a finite number of degrees of freedom, is always discrete. 
This implies that the motion
is quasi-periodic, which is just the opposite of classical chaotic motion 
which possesses a continuous frequency spectrum and exhibits statistical 
approach to equilibrium. On the other hand, the correspondence principle 
requires transition from quantum to classical mechanics for all phenomena, 
including dynamical chaos. 

For a classical integrable system the energy spectrum 
can be obtained by semi-classical quantization rules. Instead for a classical chaotic 
system the complexity of the motion 
reflects in some statistical properties of the 
spectrum. In Ref~\cite{CGV80}, by considering a billiard in a stadium, it has been 
conjectured, and numerically verified, that the level spacing distribution of a 
classical chaotic systems is described by the celebrated Wigner-Dyson distribution 
od random matrix theory (RMT) which, in the 
statistical theory of spectra, has been introduced to describe the distribution of 
spectra of systems with many degrees of freedom~\cite{Haake}.
Since these early times, the level statistics of classical chaotic systems has been 
studied in great detail, 
both theoretically and numerically, and we have now a quite 
satisfactory understanding~\cite{BGS84,Guhr98}. Yet we are very 
far from having a quantum ergodic hierarchy 
of statistical properties similar to the classical one. 

For example, it is generally 
stated that the levels spacing distribution of chaotic systems is described by 
the Wigner-Dyson distribution. 
However, this statement is quite vague for at least two reasons: 

(i) The level spacings distribution might depend on the energy range inside which the 
statistics is computed. Consider for example a prototype of chaotic systems: the stadium 
billiard. Classically, this system is chaotic no matter what energy and for any value of 
the parameter $\epsilon=a/r$, where $r$ is the radius of the circle and 
$a$ is length of the 
straight segment. However, if $\epsilon<<1$ 
the levels spacing distribution is not 
given by Wigner-Dyson up to energies which become larger and larger the smaller is 
$\epsilon$. 
The reason is due to the phenomenon of dynamical localization, which leads to 
the suppression of diffusion in angular momentum~\cite{BCL96}. 

(ii) In the classical ergodic 
theory what is the minimal statistical property which leads to the Wigner-Dyson 
distribution of quantum level spacing statistics? 
Do we need mixing behaviour or is simple ergodicity sufficient?

In this paper, we address this question 
by considering the triangle map~\cite{Prosen00}, which exhibits the same 
qualitative properties of triangular or polygonal billiards, but is
much simpler for analytical and numerical investigations.
The classical dynamics is not chaotic (even though it can become so
in a generalized version of the map, discussed later in our paper, 
where the cusps in the potential are
rounded-off) and, depending on the model parameters, is mixing, ergodic, 
quasi-ergodic~\cite{quasi-ergodic-footnote}, or quasi-integrable. Our results show that ergodicity 
is a sufficient condition to obtain spectral statistics as well as
eigenfunction properties in agreement with RMT
(see~\cite{Crt21} for recent similar results for triangle billiards),
while the quasi-ergodic case, where a single trajectory fills in the 
classical phase space extremely slowly in time~\cite{Prosen00}, exhibits a 
different behavior depending on the quantity under scrutiny.
That is, level spacing statistics is in good agreement with the Wigner-Dyson
distribution in the semi-classical limit, while there exist eigenfunctions 
localized in phase space, incompatible with the predictions of RMT.

From our results it follows 
that level spacing statistics is unable to distinguish 
between chaotic and mixing or ergodic dynamics. 
For that purpose, we can consider the quantum dynamics in the time domain
and study the out-of-time-ordered correlator (OTOC)~\cite{Larkin68,Kitaev14,Maldacena16a,Maldacena16b} or the number
of harmonics of the phase-space Wigner distribution function~\cite{Chirikov81,Gu90,Gu97,Brumer97,Brumer03,harmonics08,harmonics09,harmonics10,harmonics14}.
Indeed, for chaotic dynamics the short-time behavior of 
both quantities exhibits an exponential increase 
at a rate which is determined by the largest Lyapunov exponent of the underlying classical dynamics~\cite{Galitski17,Jiaozi20,Jiaozi21}. 
Therefore, number of harmonics and OTOC can distinguish
classically chaotic systems from those which are only mixing or ergodic.
On the other hand, these quantities cannot distinguish between 
integrable and ergodic or mixing systems. 
For a rather complete picture of the manifestations of classical ergodic 
hierarchy on quantum systems one should therefore consider 
both spectral statistics and quantities in the time domain. 




Our paper is organized as follows. in Sec.~\ref{Sect-CM}, we introduce the classical and quantum triangle map, 
then we study the statistical properties 
of the quantum triangle map in Sec.~\ref{Sect-SP}, 
while dynamical properties (OTOC and number of harmonics) 
are discussed in Sec.~\ref{Sect-DM}. 
Finally, conclusions and discussions are given in Sec.~\ref{Sect-Conclusion}.

\section{The triangle map}\label{Sect-CM}

\begin{figure}
	\includegraphics[width=1\columnwidth]{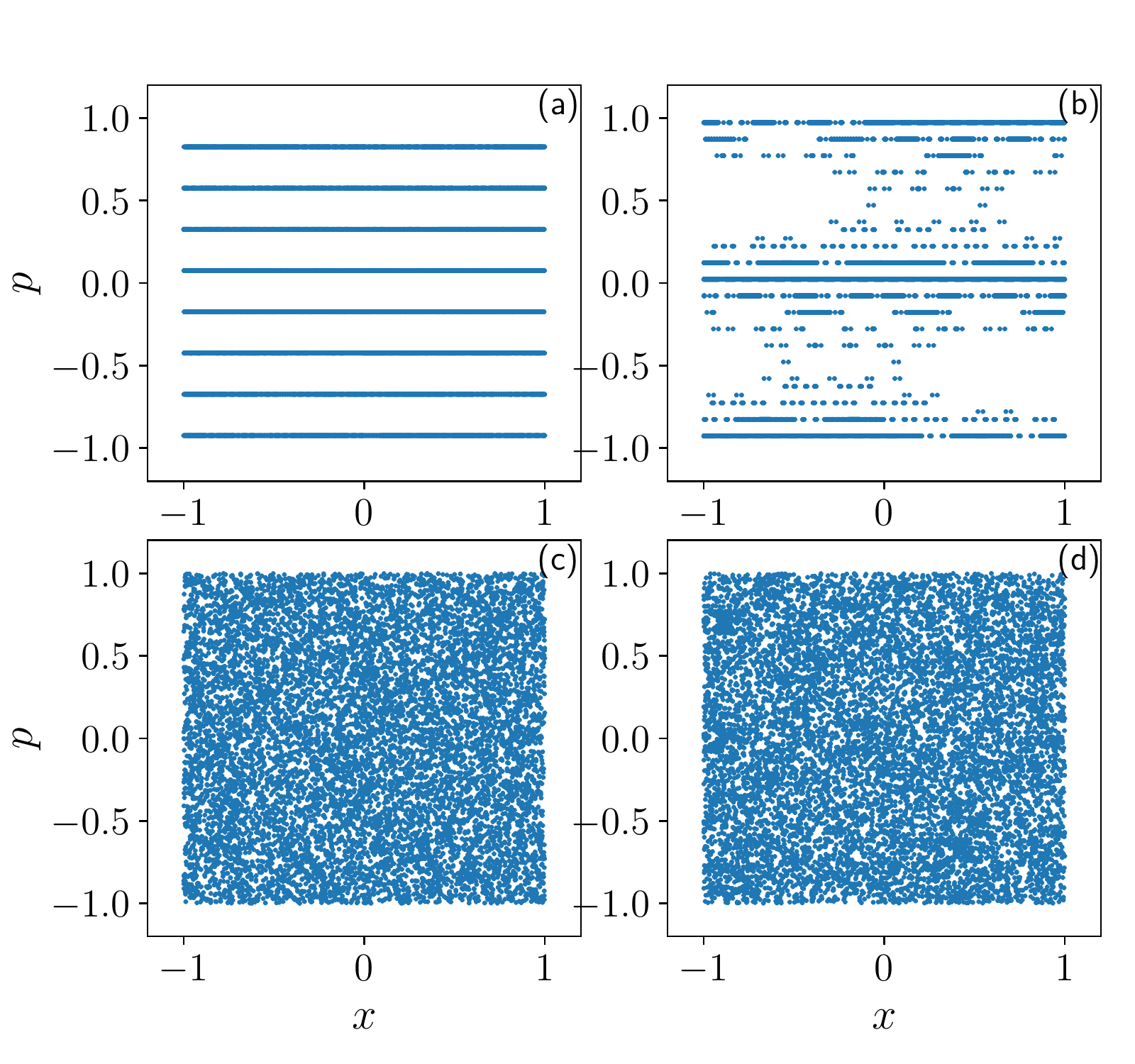}
	\caption{Poincar\'e sections 
	for the classical triangle map drawn from a single orbit up to $T = 10^4$ map iterations, for (a) pseudo-integrable case $\alpha = 1/2,\ \beta = 1/4$, (b) quasi-ergodic 
	case $\alpha = (\frac{\sqrt{5}-1}{2}-e)/2, \ \beta = 0$, (c) ergodic case $\alpha = 0, \ \beta = (\frac{\sqrt{5}-1}{2}-e)/2$, (d) ergodic and mixing case, $\alpha = (\frac{\sqrt{5}-1}{2}-e)/2,\ \beta = (\frac{\sqrt{5}-1}{2}+e)/2$.}
\label{Fig-SS}
\end{figure}

The classical triangle map~\cite{Prosen00}  is 
defined on the torus with coordinates $(x,p)\in[-1,1)\times[-1,1)$ as follows:
\be
\begin{cases}\label{eq-trmap}
	p_{n+1}=p_{n}-V'(x_n)
	& (\text{mod\,2}),\\
	x_{n+1}=x_{n}+p_{n+1} & \text{(mod\,2)},
\end{cases}
\ee
where $V(x)=-\alpha|x|-\beta x$.
It is an area preserving, parabolic, piecewise linear map, which is related to a
discrete bounce map for the billiard in a triangle. The map is marginally 
stable, i.e.,
initially close trajectories separate linearly with time. Even though
the Lyapunov exponent is zero, numerical evidence indicates that this map can exhibit very different dynamics according to the parameters $\alpha$ and $\beta$. Specifically, we shall consider four cases:  
(i) $\alpha$ and $\beta$ are incommensurable irrational numbers: the map is ergodic and mixing;
(ii) irrational $\beta$ and $\alpha = 0$: the map is known to be  ergodic but not mixing~\cite{Furstenberg};
(iii) $\alpha$ is an irrational number and $\beta =0$:
the system is quasi-ergodic (also referred to as weak ergodic~\cite{Prosen00}, as extremely long times are required for a trajectory to fill in the phase space). 
(iv) the pseudo-integrable case with rational $\alpha$ and $\beta$;
As shown by the Poincar\'e sections of Fig.~\ref{Fig-SS} (drawn from a single orbit up to $T = 10^4$ map steps),
in the pseudo-integrable case only a finite number of values of $p$ appear, 
while in the ergodic and mixing cases
a single orbit covers uniformly the surface. In the quasi-ergodic case, 
the number of different values of $p$
taken by a single orbit grows only logarithmically with the number $T$ of map iterations~\cite{Prosen00}.

In the quantum case, the evolution over one map iteration is described 
by a (unitary) Floquet operator $U$ acting on the wave function $\psi$:
$\psi_{n+1}=U\psi_n$, with
\be
U=\exp\left(-i\frac{\hat{p}^{2}}{2\hbar}\right)
\exp\left(-i\frac{V(\hat{x})}{\hbar}\right),
\ee
where $\hbar=\frac{2}{\pi D}$ is the effective Planck's constant, $D$ being the Hilbert space dimension. 

\section{Statistical properties}\label{Sect-SP}

In this section, we study the statistical properties of the quantum triangle map. We consider the four different cases, corresponding to 
pseudo-integrable, quasi-ergodic, ergodic, and mixing behavior for the underlying classical triangle map.  
In Sect.\ref{SSect-SP} we investigate the spectral statistics, 
focusing on the nearest neighbor spacing distribution, then in Sect.\ref{SSect-EF} we study the distribution of the eigenfunctions' components.


\subsection{Spectral statistics}\label{SSect-SP}

We study the standard indicator of quantum chaos, the nearest neighbor spacing distribution 
$P(s)$ of the eigenvalues of the Floquet operator $U$ of the triangle map.  

\begin{figure}
	\includegraphics[width=1\columnwidth]{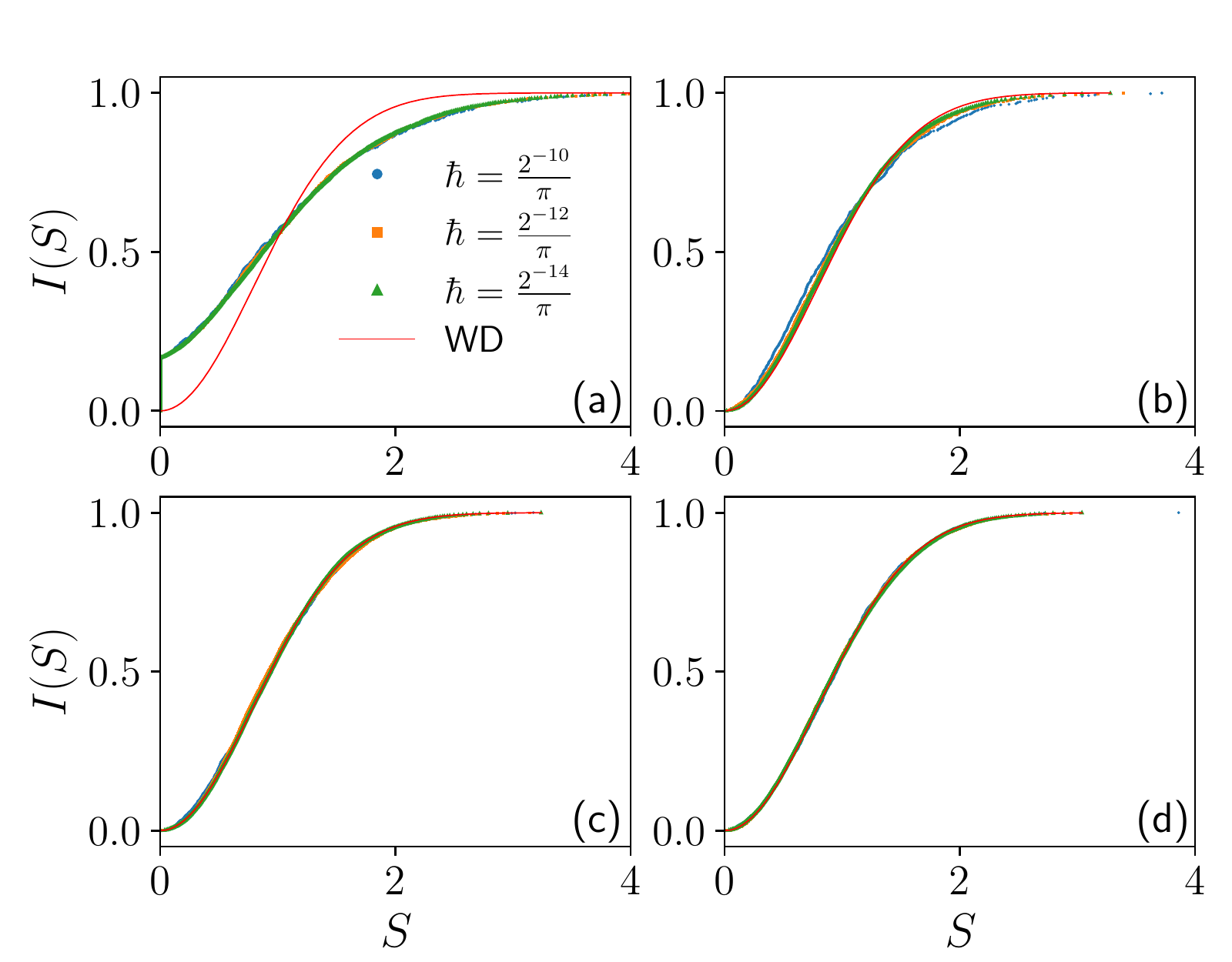}
	\caption{Cumulative distribution $I(s)$ of nearest level spacings for the triangle map,
	in the four cases (pseudo-integable, quasi-ergodic, ergodic, and mixing) 
	of Fig.~\ref{Fig-SS}. The red solid line indicates the Wigner-Dyson cumulative distribution.}
	\label{Fig-PS}
\end{figure}

\begin{figure}
	\includegraphics[width=1\columnwidth]{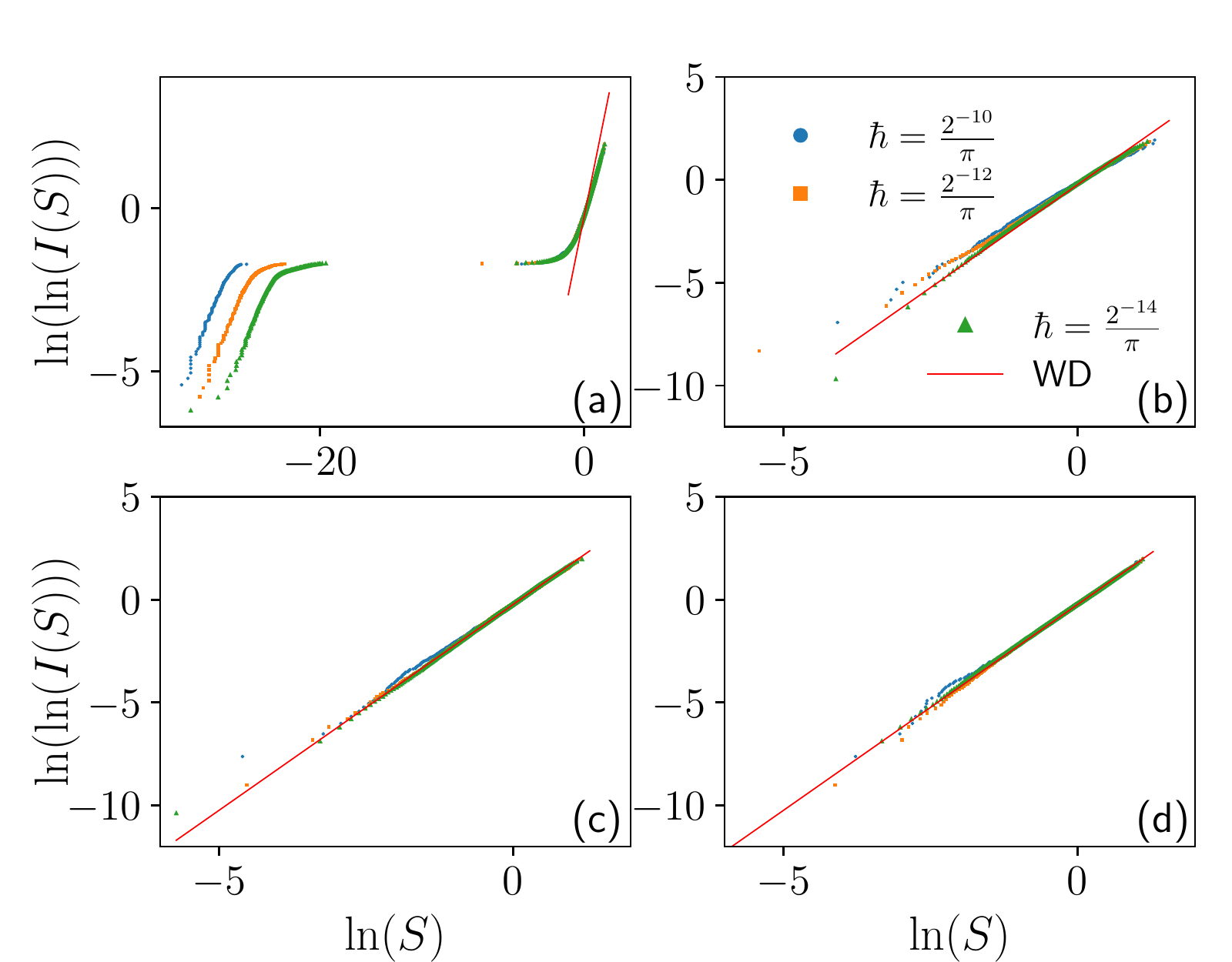}
	\caption{$\ln{(\ln{(I(s))})}$ versus $\ln{(s)}$ for four cases of Fig.~\ref{Fig-SS}. 
	The red solid line indicates the Wigner-Dyson cumulative distribution.}\label{Fig-PS-Brody}
\end{figure}

First of all, we show in Fig.~\ref{Fig-PS} the cumulative distribution $I(s)=\int_0^s ds'\,P(s')$,
and compare it 
with the predictions of Random Matrix Theory, that is, 
with the cumulative distribution obtained from the Wigner-Dyson distribution of level spacings:
\be
I_W(s) = 1 - \exp{(-\pi/4s^2)}.
\ee
In order to see clearly the distance between $I(s)$ and $I_W(s)$, 
we also show  (Fig.~\ref{Fig-PS-Brody}) 
$\ln(\ln(I(s)))$ as a function of $\ln(s)$.

\begin{figure}
	\includegraphics[width=0.8\columnwidth]{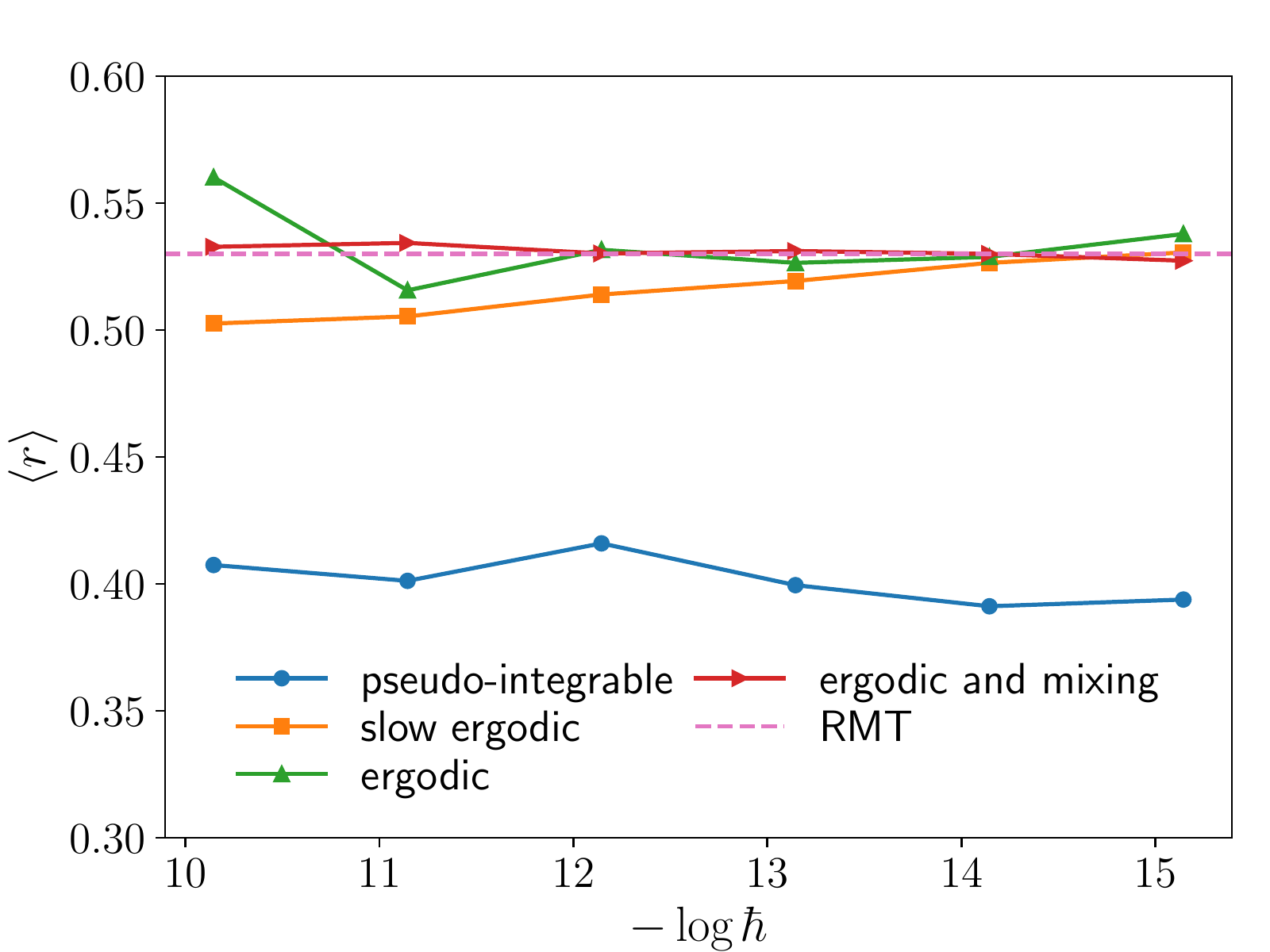}
	\caption{Parameter $\langle r \rangle $ for the four cases of Fig.~\ref{Fig-SS}. 
	The dashed line indicates the RMT prediction $\langle r \rangle\approx 0.53$.}\label{Fig-Pr}

\end{figure}

Furthermore, to see how the distance between the level spacing distribution and 
the Wigner-Dyson distribution evolves approaching the classical limit, 
we also consider 
(see Fig.~\ref{Fig-Pr}) the parameter $\langle r \rangle$~\cite{Oganesyan,Bogomolny}
as a function of the effective Planck's constant $\hbar$. 
The parameter $\langle r \rangle$ is obtained as mean value of $r_k$, defined as
\be
r_k = \frac{\min({\Delta_k, \Delta_{k+1}})}{\max{(\Delta_k,\Delta_{k+1}})},
\ee
where $\Delta_k$ is the $k$-th level spacing of the spectrum of the Floquet operator. 
This parameter characterizes the correlations between adjacent gaps in the spectrum 
and RMT predicts $\langle r \rangle \approx 0.53$. 

Our numerical simulations for all the above quantities 
show an excellent agreement with the predictions of RMT in the 
mixing case, and even in the ergodic but not mixing case. 
While the quantum chaos conjecture predicts statistical properties of classically chaotic quantum systems in agreement 
with the predictions of RMT, our results for a non-chaotic system 
suggest that the conjecture can be extended to a broader class of systems, 
with ergodicity as a sufficient condition for the onset  of RMT behavior. 

While the pseudo-integrable case clearly deviates from RMT predictions, the 
quasi-ergodic case is close to RMT, even though some
differences are visible. On the other hand,
data from Fig.~\ref{Fig-Pr} might suggest 
that spectral statistics 
approaches slowly the RMT predictions
with decreasing $\hbar$, that is, when approaching the  classical limit.

\subsection{Eigenfunction of the Floquet operator}\label{SSect-EF}

To further investigate the statistical properties of the triangle map, 
in this section we consider the distribution of the rescaled components (in the basis of the eigenstates of the operator $\hat{x}$, rescaled by $1/\sqrt{D}$) 
 of each individual 
eigenfunction of the Floquet operator.  

Since the triangle map is a time-reversal symmetric system, it is convenient to consider the symmetric Floquet operator
\be
U_{\rm sym}=\exp\left(-i\frac{V(\hat{x})}{2\hbar}\right)\exp\left(-i\frac{\hat{p}^{2}}{2\hbar}\right)\exp\left(-i\frac{V(\hat{x})}{2\hbar}\right),
\ee
so that all the components of an eigenfunctions are real (up to an irrelevant global phase factor). 

A first indicator of the agreement with the predictions of RMT for the components of an eigenfunction
is the  ratio between the second squared moment and the fourth moment
of the components distribution:
\be
\Lambda = \frac{({\cal M}_2)^2}{{\cal M}_4}.
\ee
Note that RMT predict a Gaussian distribution of the components, for which $\Lambda=1/3$.

\begin{figure}
	\includegraphics[width=1\columnwidth]{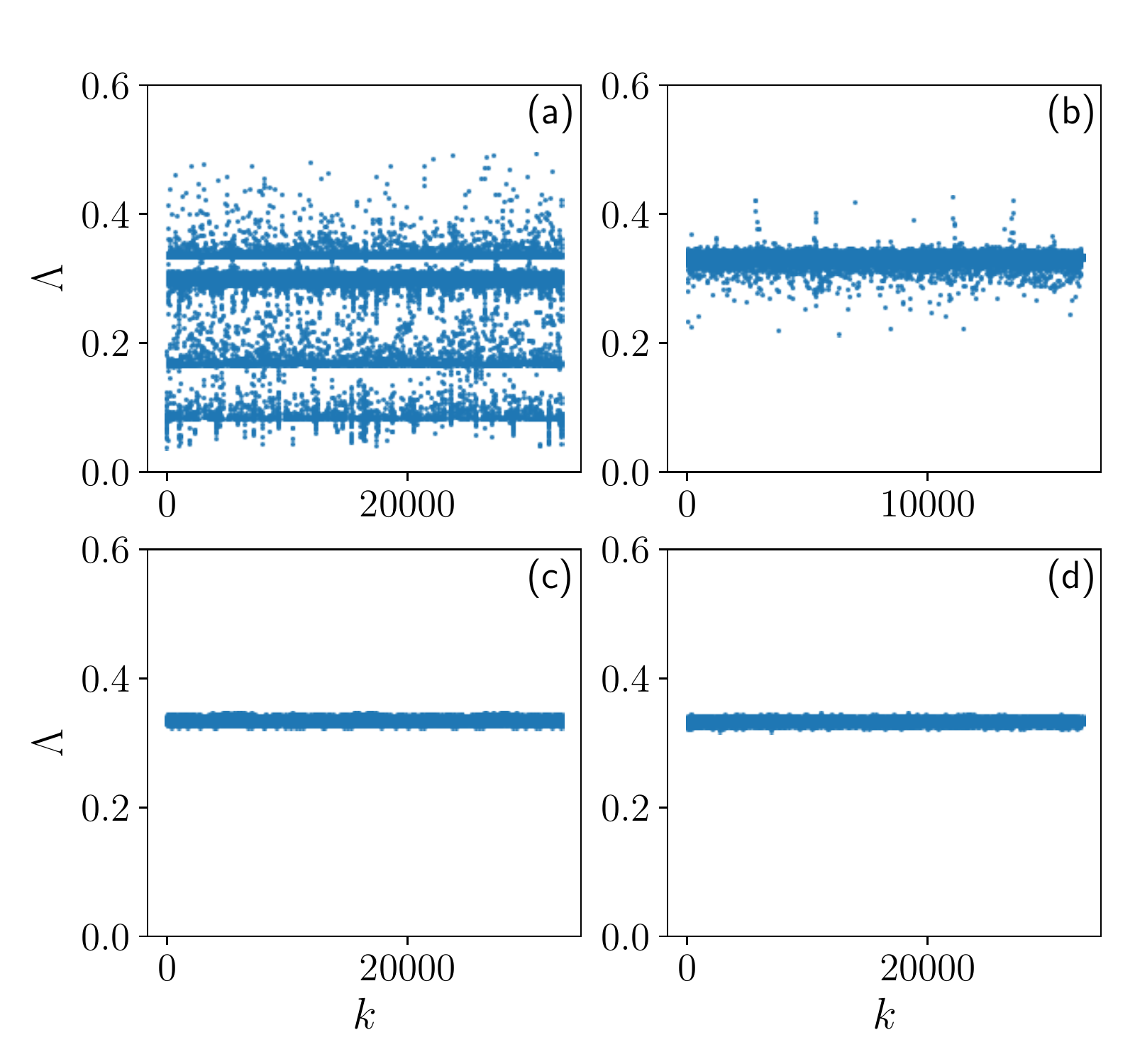}
	\caption{The ratio $\Lambda$ for every eigenstate 
	($k=1,...,D$, where $D=2^{15}$ is the Hilbert space dimension)
	in the triangle map for the four cases of Fig.~\ref{Fig-SS} and $\hbar=\frac{2^{-14}}{\pi}$.
	}\label{Fig-Lambda}
\end{figure}

\begin{figure}
	\includegraphics[width=1\columnwidth]{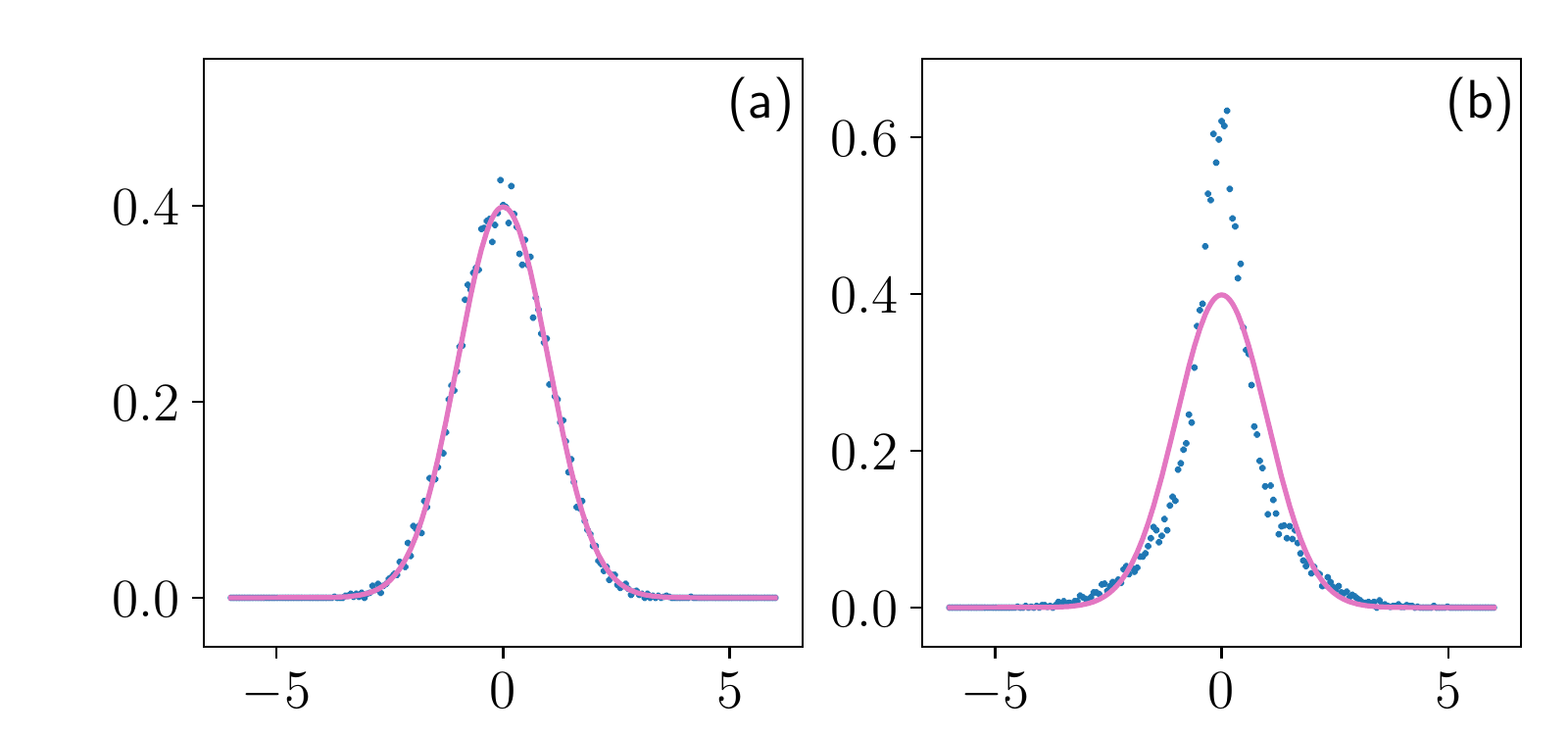}
	\includegraphics[width=0.7\columnwidth]{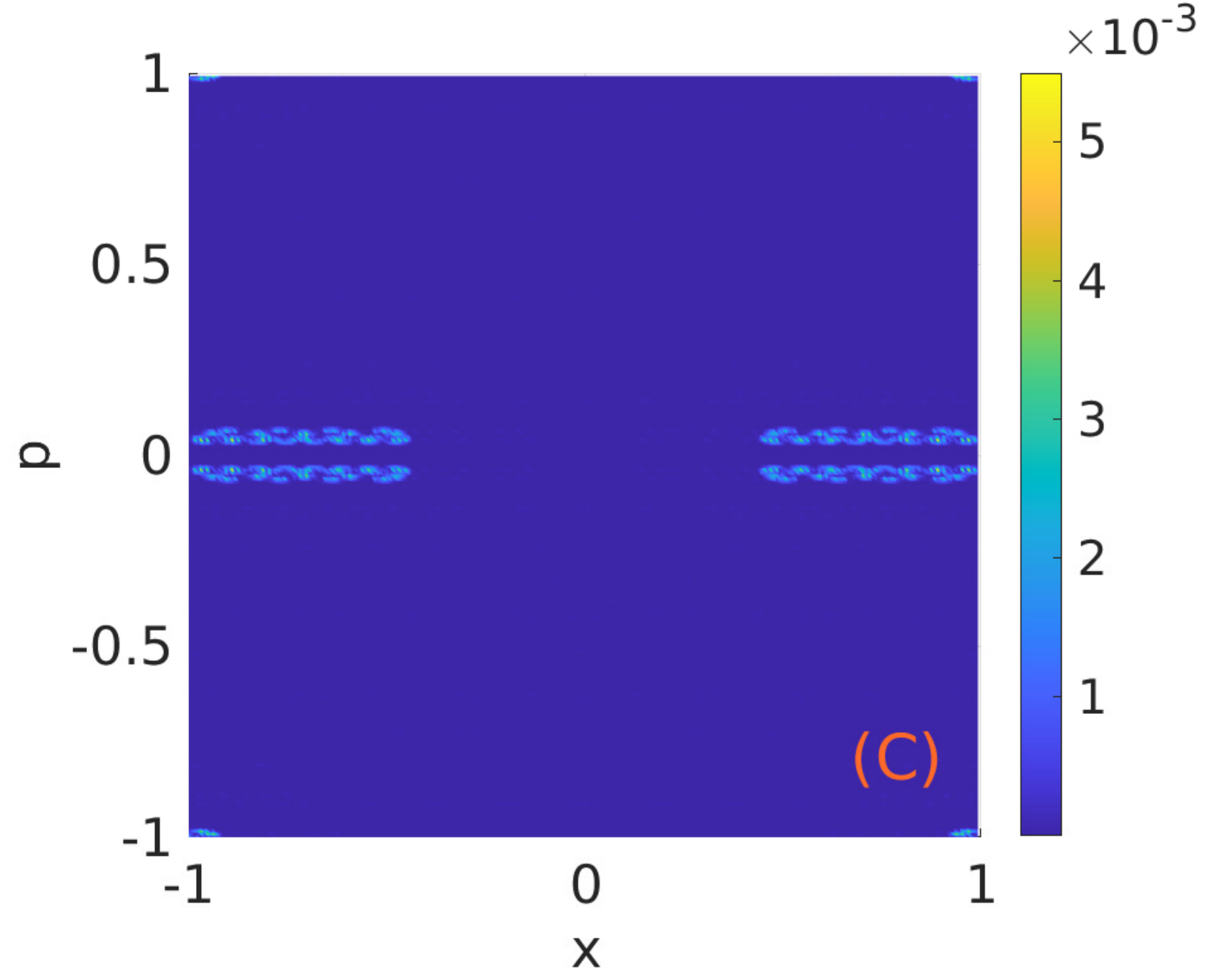}
	\vskip-0.3cm
	\caption{Distribution of rescaled components(rescaled by $1/\sqrt{D}$) of eigenfunctions, for both (a) ``extended"($\Lambda \simeq 1/3$) and (b) ``localized" (smallest value of $\Lambda$) states in the quasi-ergodic case $\alpha = (\frac{\sqrt{5}-1}{2}-e)/2, \ \beta = 0$ and $\hbar=\frac{2^{-14}}{\pi}$.; (c) Husimi distribution for the localized state shown in (b).} \label{TM-Dis}
\end{figure}

The value of the quantity $\Lambda$ for each individual state is shown in Fig.\ref{Fig-Lambda},
for $\hbar=\frac{2^{-14}}{\pi}$. On can see that both in 
the mixing and ergodic cases  $\Lambda\simeq 1/3$ holds for almost all individual states. 
This result suggests, similarly to the above study of the spectral statistics, that ergodicity in the classical case is sufficient for the 
onset of RMT in the quantum case. The pseudo-integrable case is, as expected, incompatible with the RMT predictions. 
Finally, for the quasi-ergodic case most of the eigenstates have values $\Lambda\simeq 1/3$.
On the other hand, there exist some localized states,
with the value $\Lambda$ far from $1/3$.  
The Husimi function of the state with smallest value of $\Lambda$ is shown in Fig.~\ref{TM-Dis}(c): one can see that such state is indeed localized in phase space.
The detailed distribution of rescaled components of some individual states are also shown in Fig.~\ref{TM-Dis}(a,b),
where one can see that for ``extended" states with $\Lambda\simeq 1/3$, the distribution indeed has a Gaussian shape, 
while for ``localized" state, the distribution is far from being Gaussian.

It should be stressed here that the quasi-ergodic case is a little subtle in both
classical~\cite{quasi-ergodic-footnote} and quantum cases, and deserves further investigations to be clarified, which 
is beyond the scope of our current paper. 


\section{Dynamical properties}\label{Sect-DM}

The results of  Sec.~\ref{Sect-SP} show that spectral statistics can be used to detect ergodicity but not 
chaos in the semiclassical regime. For the latter purpose, one can investigate the short-time dynamics and use
quantities such as the number of harmonics and the OTOC. As we shall discuss below, both quantities 
can measure the Lyapunov exponent, from their time-evolution within the Ehrenfest time scale.


\subsection{Out-of-time-ordered correlation}\label{SSect-OTOC}
Here we consider the averaged OTOC, defined as
\be
C_L(t)=\frac{1}{N}\sum_{k=1}^{N}\ln\left(\langle\psi_{k}|[\hat{x}(t),\hat{p}(0)]^{2}|\psi_{k}\rangle\right).
\label{eq:OTOCSchwinger}
\ee
The average is taken over a large number $N$ of initial coherent states, 
\be
\psi_{k}(x)=(\pi\hbar)^{-1/4}\exp\left(-\frac{(x-x_{k})^{2}}{2\hbar}
+\frac{ip_{k}x}{\hbar}\right),
\ee
with $(x_k,p_k)$ is the center of the $k$-th initial state,  randomly distributed in 
the phase space. 
The  classical correspondence of $C_L(t)$, denoted as $C_L^{\rm cl}(t)$,
is obtained by the canonical substitution
$\frac{1}{i\hbar}[\hat{A},\hat{B}]\rightarrow \{A,B\}_{\text{PB}}$,
where the right-hand side is the Poisson bracket of the classical 
variables $A$ and $B$.
We can write the classical counterpart of the OTOC as
\be\label{eq-altan}
C_L^{\rm cl}(t)=\frac{1}{N}\sum_{k=1}^{N}\ln\left[\int d\boldsymbol{\gamma}\rho_{\boldsymbol{\gamma}_{0}^{k}}(\boldsymbol{\gamma})\left(\frac{\partial x(t)}{\partial x(0)}\right)^{2}\right],
\ee
where $\gamma=(x,p)$. The initial condition is a Gaussian distribution
\be\label{eq:InitialC}
\rho_{\boldsymbol{\gamma_{0}^{k}}}(\boldsymbol{\gamma})=(2\pi\sigma^{2})^{-1}
\exp\left(-\frac{(x-x_{k})^{2}+(p-p_{k})^{2}}{2\sigma^{2}}\right),
\ee
where, in order to compare with the quantum wavepacket, 
we take $\sigma=\sqrt{\frac{\hbar}{2}}$.

Note that we take the average of the logarithms of the OTOC of the individual initial states $\psi_{k}$.
The simple average, instead, would lead to a divergence since there are
singular points (the cusps in the potential) for which $\partial x(t)/\partial x(0)$ diverges. 
This is a general problems when taking the simple average of 
OTOCs~\cite{Hirsch19,Cao20,Bunimovich20}, as it is dominated in integrable systems by 
the local instability of unstable fixed points, which might lead to an exponential growth completely unrelated 
to chaos. On the other hand, as we shown in Ref.~\cite{Jiaozi21}, a proper averaging (of the logarithms of OTOC) restores the role of OTOC
as a diagnostic of chaotic dynamics, as
\be
C_L(t)\propto 2\lambda t,
\ee
where $\lambda$ is the largest Lyapunov exponent of the underlying dynamics, and this growth lasts up to the 
Ehrenfest time scale. Therefore, we expect a linear in time growth of $C_L(t)$ for chaotic dynamics, with the growth rate 
measuring the ``quantum Lyapunov'' exponent.

\begin{figure}
	\includegraphics[width=1\columnwidth]{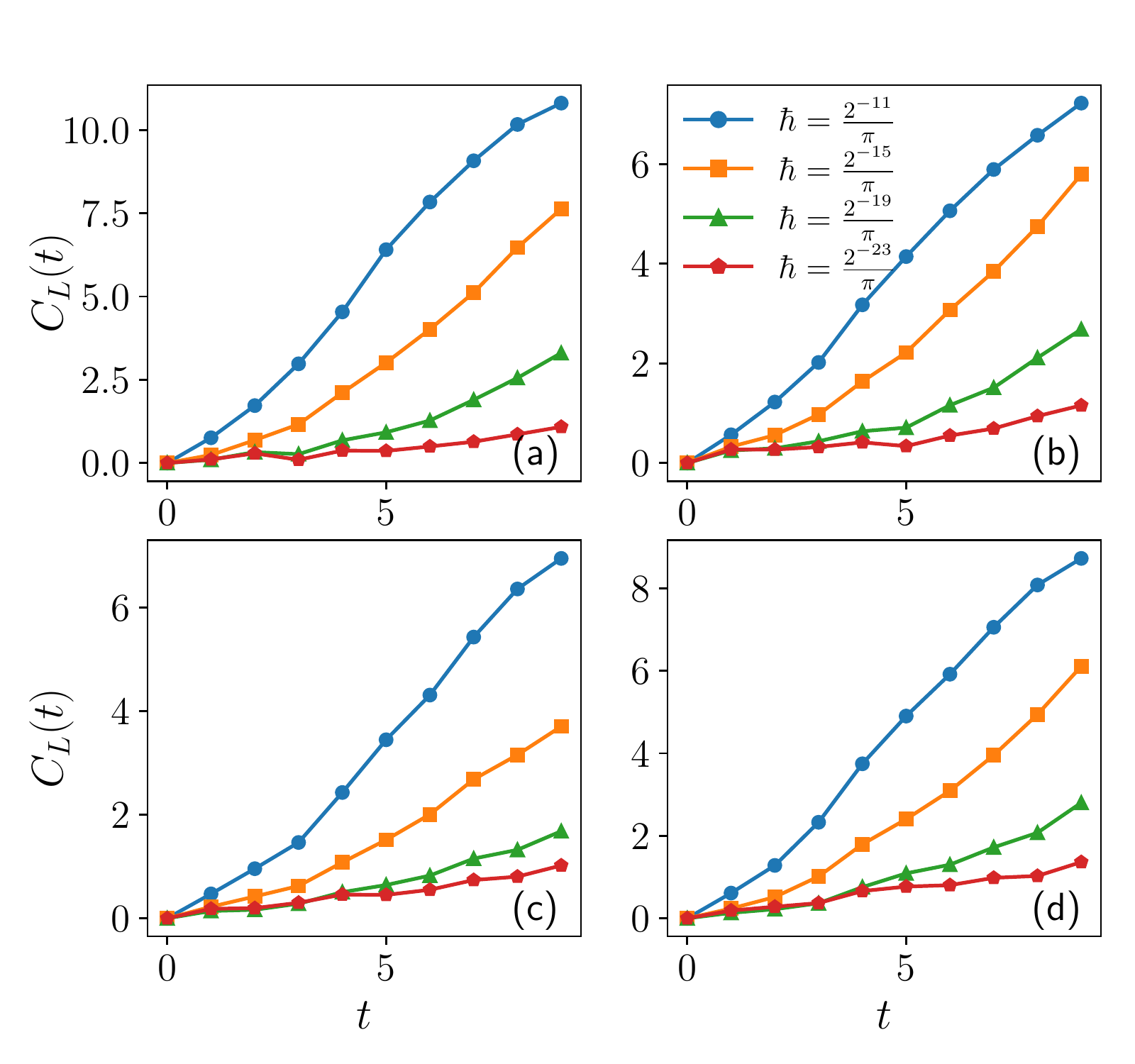}
	\caption{The averaged OTOC $C_L(t)$ in triangle map for the four different cases of 
	Fig.~\ref{Fig-SS}.}\label{Fig-OTOC}
\end{figure}

It can be seen from Fig.~\ref{Fig-OTOC} that, for all the cases above considered (pseudo-integrable, quasi-ergodic, ergodic, and mixing), 
$C_L(t)$ behaves similarly. That is, for $\hbar\to 0$ the growth rate vanishes, 
indicating a zero ``quantum Lyapunov" exponent in the semiclassical limit.

\subsection{Number of harmonics}\label{SSect-M2}
For classical chaotic dynamics, the exponential sensitivity
implies that the density distribution 
in the phase space is exponentially stretched and folded, 
becoming increasingly intricate on smaller and smaller scales. 
Therefore, in order to reconstruct 
the increasingly finer details of 
the phase-space distribution,  
the number of harmonics, that is, components 
in Fourier space, also increases 
exponentially in time~\cite{Gu90,Brumer97,Brumer03}.
In an integrable system this does not happen, as the instability 
is typically linear in time~\cite{Casati80}.
It follows that the growth rate of the number of harmonics of the phase-space distribution
can be used, similarly to the Lyapunov exponent, as a way 
to characterize classical chaos. The advantage is that this phase space approach,
in contrast with the exponential instability of trajectories, can be transferred to the 
quantum domain. 

Indeed, we consider the Fourier transform of the Wigner function $W(x,p)$:
\begin{equation}
W(x,p)=\sum_{m,n} \tilde{W}_{m,n}e^{-i\pi (mx+np)}.
\end{equation}
The number of harmonics is then  estimated from the square root of the second 
moment ${\cal M}_{2}$ of the harmonics distribution:
\begin{equation}
{\cal M}_{2}= \sum_{m,n} (m^2+n^2) \tilde{W}_{m,n}.
\end{equation}
For pure initial states, we obtain~\cite{Brumer03}
\begin{equation}
{\cal M}_{2}= \frac{2}{\hbar^2} \,(\Delta_2(\hat{x})+\Delta_2(\hat{p})),
\end{equation}
where
\begin{equation}
\Delta_2(x)=\langle \hat{x}^2 \rangle - \langle \hat{x}\rangle^2,
\quad
\Delta_2(p)=\langle \hat{p}^2 \rangle - \langle \hat{p}\rangle^2.
\end{equation}

\begin{figure}
	\includegraphics[width=1\columnwidth]{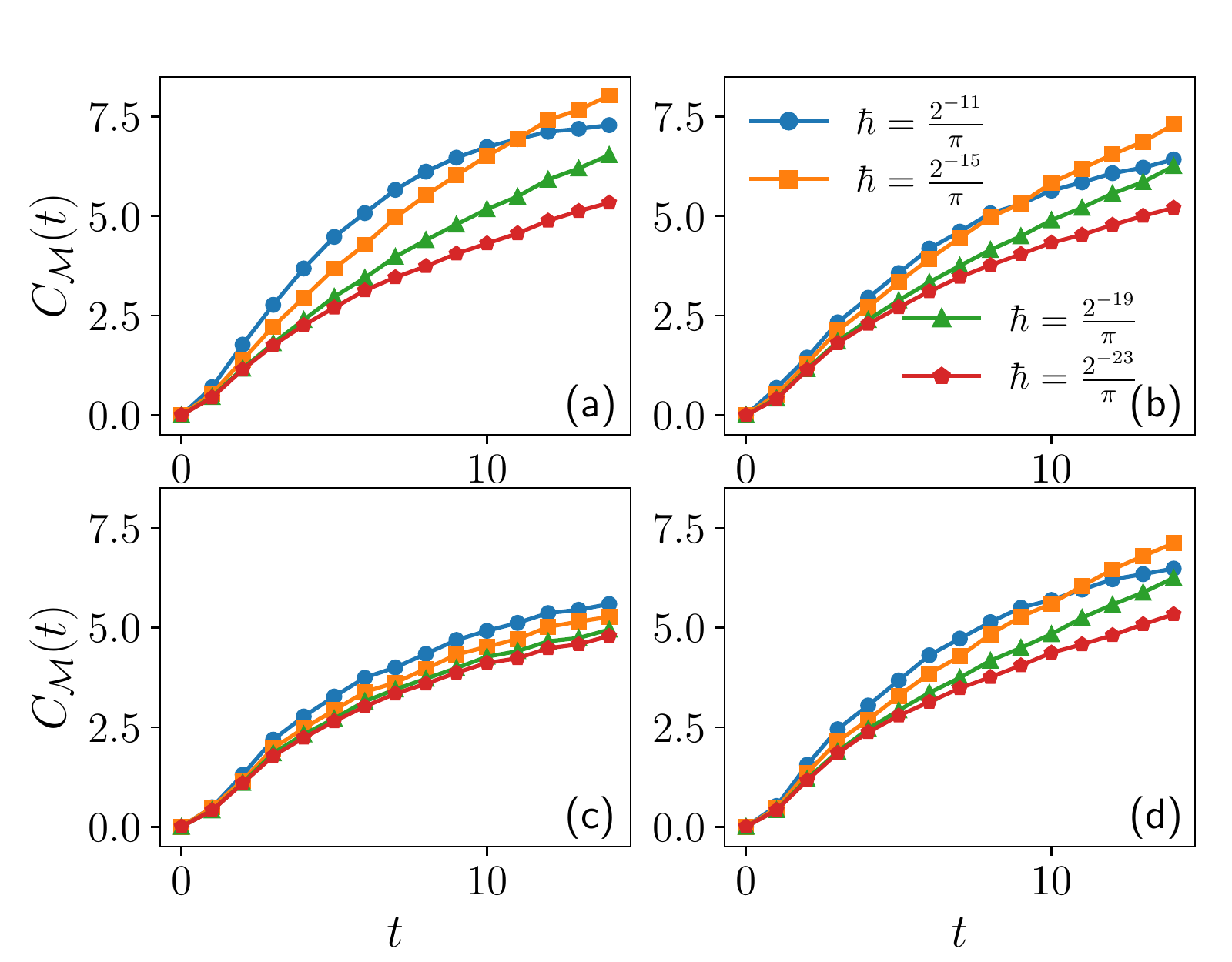}
	\caption{The averaged number of harmonics $C_{\cal M}(t)$ in triangle map for the 
	four different cases of Fig.~\ref{Fig-SS}.}
	\label{fig:harmonics}
\end{figure}

We then average 
the logarithms of the second moment 
${\cal M}_{2}^{(k)}(t)$ for the initial states $\{\psi_k\}$: 
\be
C_{{\cal M}}(t)=\frac{1}{N}\sum_{k=1}^{N}\ln\left( {\cal M}_{2}^{(k)}(t)\right),
\label{eq:M2}
\ee
where the average is performed as in the case of OTOC. 

The numerical results, shown in Fig.~\ref{fig:harmonics}, confirm the main message obtained from the study of the OTOC. 
We cannot find an $\hbar$-independent growth rate
for the number of harmonics, in all the considered cases (pseudo-integrable, quasi-ergodic, ergodic, and mixing).
These results should be contrasted with those of 
the chaotic case, where a growth rate given by twice the Lyapunov 
exponent is observed 
(see discussion in Sec.~\ref{subsec:chaotic} and the data
shown in Fig.~\ref{TMR-All}).

\subsection{Comparison with chaotic case}
\label{subsec:chaotic}

\begin{figure}
	\includegraphics[width=1\columnwidth]{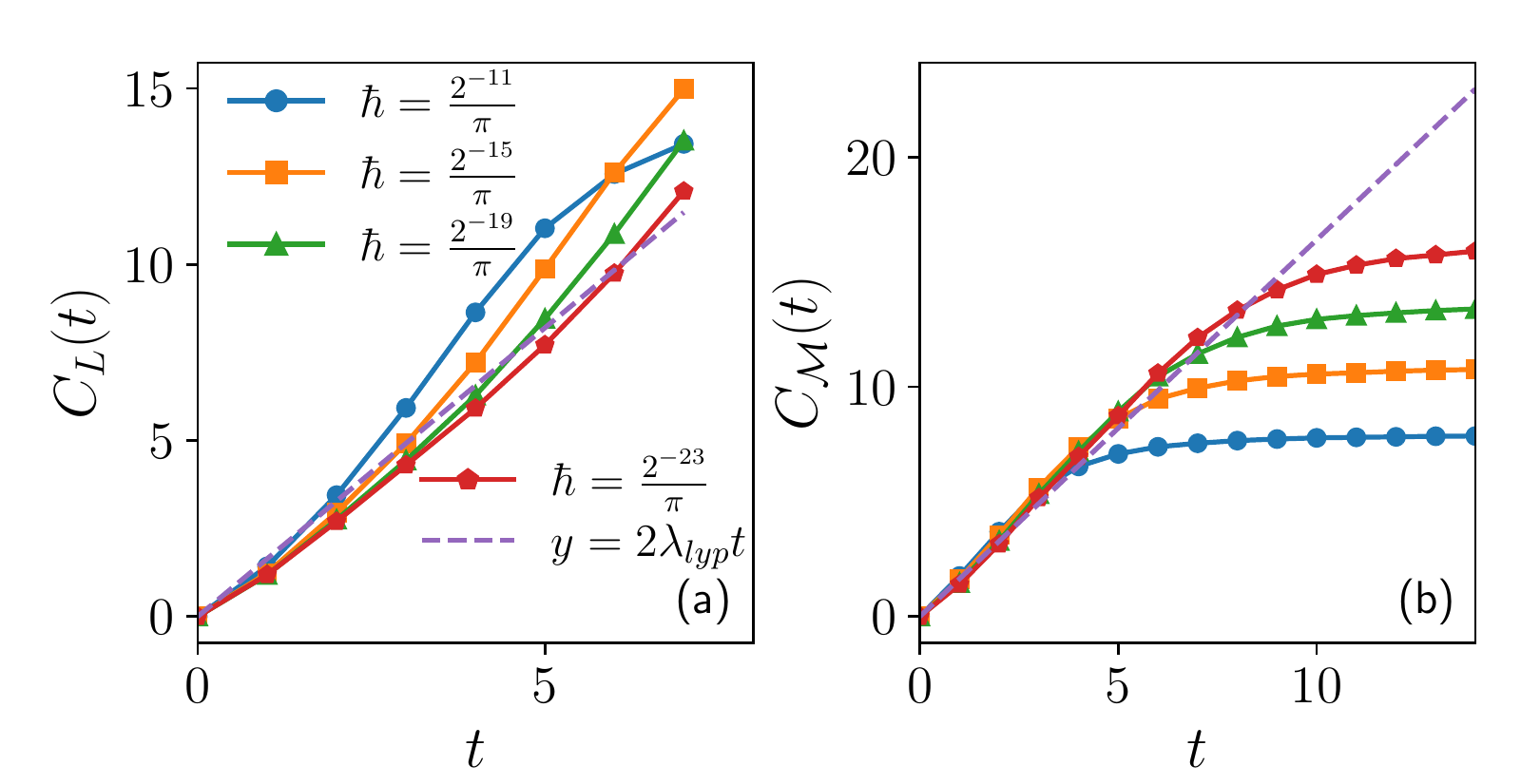}
	\caption{The averaged OTOC $C_L(t)$ and the averaged number of harmonics $C_{\cal M}(t)$ in the round-off triangle map for $r=0.2$, $\alpha = (\frac{\sqrt{5}-1}{2}-e)/2, \ \beta = 0$.
	Note the sharp contrast between this figure and 
	Figs.~\ref{Fig-OTOC} and \ref{fig:harmonics}: in those 
	figures the growth rate decreases with $\hbar$, while 
	here it is $\hbar$-independent.}\label{TMR-All}
\end{figure}

We compare the results obtained so far with those for a chaotic system. 
We consider a round-off triangle map~\cite{Jiaozi21}, in which
we substitute the cusps in the potential 
$V(x)$ of the original triangle map by small circle arcs of radius $r$:
\be\label{eq-RTM-Vx}
\frac{V(x)}{\alpha}=\begin{cases}
        -\sqrt{2}r+\sqrt{r^{2}-x^{2}}& |x|\le\frac{\sqrt{2}}{2}r,\\
        -1+\sqrt{2}r-\sqrt{r^{2}-(|x|-1)^{2}}&|x|\ge 1-\frac{\sqrt{2}}{2}r,\\
        -|x|&{\rm otherwise}.
\end{cases}
\ee
Differently from the 
original triangle map (recovered for $r=0$), the round-off triangle map is exponentially unstable for any $r\ne 0$,
that is,  the maximum Lyapunov exponent is positive. In Fig.~\ref{TMR-All}, 
we show OTOC and number of harmonics for $r = 0.2$. One can see 
that for both quantities the initial growth follows, up to the Ehrenfest time,  
a slope given by twice the Lyapunov exponent of the classical chaotic dynamics. 
Therefore, OTOC and number of harmonics can be used to detect and quantify 
exponential instability in the semi-classical limit $\hbar\rightarrow 0$. 
This is not possible using spectral statistics, which follows Random Matrix Theory
for chaotic but also for non-chaotic, mixing or even uniquely ergodic systems.

\section{Conclusions and Discussions}\label{Sect-Conclusion}

By studying the statistical properties of different cases of the quantum triangle map, we find that ergodicity is a sufficient condition for the onset of random matrix behavior, in the properties of both spectrum and eigenfunctions.
Therefore, to detect classically chaotic motion in a quantum system, one should
rather consider quantities in the time domain, such as the OTOC and the number 
of harmonics, whose growth rate is determined by the classical Lyapunov exponent. 
While such quantities are sensitive indicators of classical chaotic dynamics, 
in order to distinguish between integrable and ergodic or mixing dynamics 
one should instead use spectral statistics. 
Given the quantities discussed in this paper, the problem remains to distinguish between mixing and uniquely ergodic dynamics. This problem could be tackled 
by looking at the decay of correlation functions, as done for the classical
triangle map~\cite{Prosen00}.

{\it Acknowledgments}:
J. Wang acknowledges the financial support of the Deutsche Forschungsgemeinschaft (DFG), Grants No. 397107022 (GE 1657/3-2) and No. 397067869 (STE 2243/3-2) within the DFG Research Unit FOR 2692.  G.B. acknowledges the financial support of the INFN through the project “QUANTUM”. W.-g. Wang acknowledges the financial support of the Natural Science Foundation of China under Grant
 Nos.~11535011, 11775210, and 12175222.



\begin{thebibliography}{99}

\bibitem{Haake}
F. Haake, S. Gnutzmann, and M. Ku\'{s},
\textit{Quantum Signatures of Chaos} (4th ed.)
(Springer, 2018).

\bibitem{CGV80}
 G. Casati, F. Valz-Gris, and I. Guarneri,
 Lettere al Nuovo Cimento \textbf{28}, 279 (1980).
 
\bibitem{BGS84}
O. Bohigas, M.-J. Giannoni, and C. Schmit, 
Phys. Rev. Lett. \textbf{52}, 1 (1984).
In this paper, the model studied in Ref~\cite{CGV80} 
was investigated in a more complete and clear fashion 
and the conjecture in~\cite{CGV80} confirmed.
 
\bibitem{Guhr98}
T. Guhr, A. M\"{u}ller-Groeling, H.A. Weidenm\"{u}ller, 
Phys. Rep. \textbf{299}, 189 (1998). 

\bibitem{BCL96}
F. Borgonovi, G. Casati, and B. Li,
Phys. Rev. Lett. \textbf{77}, 4744 (1996). 

\bibitem{Prosen00} 
G. Casati and T. Prosen, Phys. Rev. Lett. \textbf{85}, 4261 (2000).

\bibitem{quasi-ergodic-footnote}
Quasi-ergodicity is a weaker property than ergodicity.
Indeed, for quasi-ergodic systems one cannot prove that 
time averages are equal to phase averages.
For instance, for the system of two elastic hard-point masses 
in one dimension with generic mass ratio, numerical results 
indicate that the orbits are dense on the energy surface, yet 
the system is nonergodic~\cite{Jiao14}.
It is an open problem whether the triangle map 
in the quasi-ergodic regime can exhibit 
the same phenomenon.

\bibitem{Jiao14}
J. Wang, G. Casati, and T. Prosen,
Phys. Rev. E \textbf{89}, 042918 (2014).

\bibitem{Crt21}
C. Lozej, G. Casati, and T. Prosen,
preprint arXiv:2110.04168 [nlin.CD].

\bibitem{Larkin68}A. Larkin and Y. N. Ovchinnikov, Zh. Eksp. Teor. Fiz. Sov. Phys. \textbf{55}, 1200 (1968) [JETP \textbf{28}, 1200 (1969)].

\bibitem{Kitaev14} A. Kitaev, \textit{Hidden correlations in the Hawking radiation and thermal noise}, talk given at KITP, Santa Barbara, 2014, 
http://online.kitp.ucsb.edu/online/joint98/kitaev/.

\bibitem{Maldacena16a} J. Maldacena and D. Stanford, 
Phys. Rev. D \textbf{94}, 106002 (2016).

\bibitem{Maldacena16b} 
J. Maldacena, S. H. Shenker, and D. Stanford, 
J. High Energy Phys. {08} (2016) 106.

\bibitem{Chirikov81}
B. V. Chirikov, F. M. Izrailev, and D. L. Shepelyansky, Sov. Sci. Rev. C \textbf{2}, 209 (1981).

\bibitem{Gu90}  Y. Gu, Phys. Lett. A \textbf{149}, 95 (1990).

\bibitem{Gu97}
Y. Gu and J. Wang, Phys. Lett. A \textbf{229}, 208 (1997).

\bibitem{Brumer97} A. K. Pattanayak and P. Brumer, Phys. Rev. E \textbf{56}, 
5174 (1997).

\bibitem{Brumer03} J. Gong and P. Brumer, Phys. Rev. A \textbf{68}, 062103
(2003).

\bibitem{harmonics08} V. V. Sokolov, O. V. Zhirov, G. Benenti, G. Casati, 
Phys. Rev. E \textbf{78}, 046212 (2008).

\bibitem{harmonics09} G. Benenti and G. Casati, Phys. Rev. E \textbf{79}, 
025201 (R) (2009). 

\bibitem{harmonics10} V. Balachandran, G. Benenti, G. Casati, and J. Gong, 
Phys. Rev. E \textbf{82}, 046216 (2010).

\bibitem{harmonics14} P. Qin, W. Wang, G. Benenti, and G. Casati, 
Phys. Rev. E \textbf{89}, 032120 (2014).

\bibitem{Galitski17} E. B. Rozenbaum, S. Ganeshan, and V. Galitski, Phys. Rev. Lett. \textbf{118}, 086801 (2017).

\bibitem{Jiaozi20}
J. Wang, G. Benenti, G. Casati, and W. Wang,  
Phys. Rev. Res. 2, 043178 (2020). 

\bibitem{Jiaozi21}
J. Wang, G. Benenti, G. Casati and W. Wang, 
Phys. Rev. E \textbf{103}, L030201 (2021).

\bibitem{Furstenberg}
H. Furstenberg, Am. J. Math. \textbf{83}, 573 (1961).

\bibitem{Oganesyan}
V. Oganesyan and D. A. Huse,
Phys. Rev. B \textbf{75}, 155111 (2007). 

\bibitem{Bogomolny}
Y. Y. Atas, E. Bogomolny, O. Giraud, and G. Roux
Phys. Rev. Lett. \textbf{110}, 084101 (2013). 

\bibitem{Hirsch19}
S. Pilatowsky-Cameo, J. Ch\'{a}vez-Carlos, M. A. Bastarrachea-Magnani,
P. Str\'{a}nsk\'{y}, S. Lerma-Hern\'{a}ndez, L. F. Santos, and J. G. Hirsch,
Phys. Rev. E \textbf{101}, 010202 (2020).

\bibitem{Cao20}
T.  Xu, T. Scaffidi, and X. Cao, Phys. Rev. Lett. \textbf{124},
140602 (2020).

\bibitem{Bunimovich20}
E. B. Rozenbaum, L. A. Bunimovich, and V. Galitski, Phys. Rev.
Lett. \textbf{125}, 014101 (2020).

\bibitem{Casati80}
G. Casati, J. Ford, and B. V. Chirikov, Phys. Lett. \textbf{77A}, 91 (1980).




\end{thebibliography}
\end{document}